\documentstyle[12pt,cite]{article}

\begin{document}

\begin{flushright}
{\bf IFUSP/P-1008\\ hep-th/9209086 \\
September, 1992}
\end{flushright}

\vspace{1cm}

\begin{center}
{\huge{\bf Quantization of  Spinning Particle
 with Anomalous Magnetic Momentum}}

\vspace{1cm}

{\large{\bf D.M.Gitman {\it and} A.V.Saa}}

\vspace{.5cm}

{\it Instituto de F\'{\i}sica \\
Departamento de F\'{\i}sica Matem\'atica \\
Universidade de S\~ao Paulo, Caixa Postal 20516 \\
01498-970 S\~ao Paulo, S.P., Brazil\\}

\vspace{1cm}

{\bf Abstract}
\end{center}
{\small
A generalization of the pseudoclassical action of a spinning
particle in
the presence of an anomalous magnetic moment is given. The leading
considerations, to write the action, are gotten from the path integral
representation for the causal Green's function of the generalized
(by Pauli) Dirac equation for the particle with anomalous magnetic momentum
in an external electromagnetic field. The action can be written in
reparametrization and supergauge invariant form. Both operator (Dirac) and
path-integral (BFV) quantization are discussed. The first one leads to the
Dirac-Pauli equation, whereas the second one gives
the corresponding propagator.
One of the nontrivial points  in this case is that both  quantizations
schemes demand for consistency to take into account an operators ordering
problem.}

\newpage

\section{Introduction}

In recent years numerous classical models of relativistic
particles and superparticles  were discussed
intensively.
First, the interest to such models was initiated by the
close relation
with the string theory, but now it is also clear that the problem itself
has an important meaning for the
 deeper understanding of the structure of quantum
theory.

One of the basic, in the above mentioned set of classical models, is the
pseudoclassical model of Fermi particle with  spin $1/2$, proposed first
in the works \cite{ber,casal}, investigated and quantized in many
works, see for example
 [\citen{ber,casal,brink,vechia,hen,sund,mann,gittyu,gitfrad}].
The model can be formulated in gauge
invariant (reparametrization and supersymmetric) form.
The action of the model in an external electromagnetic
field has the form \cite{gitfrad}:
\begin{eqnarray}
 S &=& \int_0^1 \left[
 -\frac{\dot{x}^2}{2e}  - e\frac{m^2}{2} - g\dot{x}^\alpha A_\alpha
+igeF_{\alpha\beta}\psi^\alpha\psi^\beta \right. \nonumber \\
 &+&\left. i\left(\frac{\dot{x}_\alpha\psi^\alpha}{e} -
m\psi^5
\right)\chi -i\psi_n\dot{\psi}^n
\right]d\,\tau\, ,
\label{lag}
\end{eqnarray}
where $x^\alpha$, $e$ are even and $\psi^n$, $\chi$ are odd variables dependent
on $\tau$, the latter plays the role of the time in this theory,
$A_\alpha(x)$ is an external electromagnetic field potential,
 $F_{\alpha\beta}(x)$
is the Maxwell strength tensor, and g the electrical charge. Greek indices
run over $\overline{0,3}$ and Latin indices $n,m$ run over $\overline{0,3},5$.
The metric tensors:
$\eta_{\alpha\beta}= {\rm diag}(1,-1,-1,-1)$ and
$\eta_{mn}= {\rm diag}(1,-1,-1,-1,-1)$.
 There are two gauge
transformations in the theory with the action (\ref{lag}),
re\-pa\-ra\-me\-tri\-za\-ti\-ons,
\begin{equation}
\delta x = \dot{x}\xi \,\,,\,\,\,\,\,\,
\delta e = \frac{d}{dt}(e\xi) \,\,,\,\,\,\,\,\,
\delta \psi^n = \dot{\psi}^n\xi \, \,,\,\,\,\,\,\,
\delta \chi = \frac{d}{dt}(\chi\xi)\, ,
\label{re}
\end{equation}
and supertransformations,
\begin{eqnarray}
&&\delta x^\alpha = i\psi^\alpha\epsilon \,\,,\,\,\,\,\,\,
\delta e = i\chi \epsilon \,\,,\,\,\,\,\,\,
\delta \chi = \dot{\epsilon} \,\,,\,\,\,\,\,\,
\delta\psi^\alpha = \frac{1}{2e}(\dot{x}^\alpha
+ i\chi\psi^\alpha)\epsilon \,\, ,  \nonumber \\
&&\delta\psi^5 = \left[\frac{m}{2}
-\frac{i}{me}\psi^5\left(
\dot{\psi}^5
-\frac{m}{2}\chi
\right)
\right]\epsilon\,\,,
\label{su}
\end{eqnarray}
where $\xi$ are even and $\epsilon$ odd $\tau$-dependent parameters.
The spinning degrees
of freedom in such a model are described by Grassmannian
 variables, that's why the model is called pseudoclassical.
The quantization of the model in different ways  leads to the
quantum mechanics of the Dirac particle, is very instructive and
creates many useful analogies with problems of quantization of gauge
field theories.

In this work  a generalization of the
model,  when  an anomalous magnetic momentum of the particle is present, is
discussed.
The relativistic quantum theory of a spinning particle, which has both the
``normal'' magnetic momentum $g/2m$ and an ``anomalous'' magnetic
momentum $\mu$, was formulated by Pauli \cite{pauli}. In this case he
generalized the Dirac equation to the following form:
\begin{equation}
 \left( \hat{{\cal P}}_\nu \gamma^\nu - m -
\frac{\mu}{2}\sigma^{\alpha\beta}
F_{\alpha\beta} \right)\Psi(x)= 0,
\label{gendirac}
\end{equation}
where
 $\hat{{\cal P}}_\nu = i \partial_\nu - g A_\nu(x)$,
$\sigma^{\alpha\beta}=
\frac{i}{2}[\gamma^\alpha,\gamma^\beta]_-\,$,
$\,[\gamma^\alpha,\gamma^\beta ]_+=2\eta^{\alpha\beta}$,
and the notations $[A,B]_{\pm} = AB \pm BA$ are used.

\noindent
We present a generalization  of the action (\ref{lag}), whose quantization
gives the Dirac-Pauli theory. The work is organized as follow. First we
construct
a path-integral representation for the causal Green's function of the
Dirac-Pauli
equation (\ref{gendirac}). The form of this representation allows one to
identify some terms of the effective action with a classical gauge
invariant action of a spinning particle with anomalous magnetic momentum.
We find gauge and supergauge transformations of the action, analyze Lagrangian
and Hamiltonian structure of the theory.
Both operator (Dirac) and path-integral (BFV) quantizations are discussed.
 The first one leads
to the Dirac-Pauli equation, whereas the second one gives the corresponding
propagator.
One of the nontrivial points  in this case is that both  quantizations
schemes demand for consistency to take into account an operators ordering
problem.

\section{Path Integral Representation for Causal Green's
Function of Dirac-Pauli Equation}

In this section we are going to write the path integral representation for
the causal Green's function $S^{\rm c}(x,y)$ of the equation (\ref{gendirac}).
To get the result in supersymmetric form one needs to work with the transformed
by \break  $\gamma^5  = \gamma^0\gamma^1\gamma^2\gamma^3\,\,$,
$\left(\gamma^5\right)^2=-1\,$, function
$\tilde{S}^{\rm c}(x,y) = S^{\rm c}(x,y)\gamma^5\,\,\,$,
which obeys the equation
\begin{equation}
 \left( \hat{{\cal P}}_\nu \tilde{\gamma}^\nu - m\gamma^5 -
\frac{\mu}{2}\gamma^5 \tilde{\sigma}^{\alpha\beta}F_{\alpha\beta} \right)
\tilde{S}^{\rm c}(x,y)= \delta^4(x-y),
\label{grennmod}
\end{equation}
where $\tilde{\gamma}^\mu = \gamma^5\gamma^\mu$,
$\tilde{\sigma}^{\mu\nu}=\frac{i}{2}\left[
\tilde{\gamma}^\mu,\tilde{\gamma}^\nu\right]_-$.

\noindent
The matrices $\tilde{\gamma}^\nu$,  obey the same commutation relations as
initial ones $\gamma^\nu$,
$\left[\tilde{\gamma}^\mu,\tilde{\gamma}^\nu\right]_+=2\eta^{\mu\nu}$, so that
the tilde sign will be omitted hereafter. For all the $\gamma$-matrices we
have $[\gamma^m,\gamma^n]_+ = 2\eta^{mn}$.

Similar to  Schwinger \cite{sch2} we present
$\tilde{S}^{\rm c}_{\alpha\beta}(x,y)$
as a matrix element of an operator $\tilde{S}^{\rm c}_{\alpha\beta}$, but,
 in the coordinate space only,
\begin{equation}
\tilde{S}^{\rm c}_{\alpha\beta}(x,y)
= \langle x | \tilde{S}^{\rm c}_{\alpha\beta} | y \rangle,
\label{prop}
\end{equation}
where spinor indices are written explicitly for clarity only once
and will be omitted hereafter;
$| x \rangle$ are eigenvectors for some hermitian
operators of coordinates
$X^\mu$; the corresponding canonically conjugated operators of momenta
are $P_\mu$, so that:
\begin{eqnarray}
&X^\mu|x\rangle = x^\mu |x\rangle\,, \,\,\,\,\,\,\,\,
 \langle x | y \rangle = \delta^4(x-y)\,, \,\,\,\,\,\,\,\,
\int|x\rangle\langle x|dx = I\,,&   \nonumber \\
& \left[P_\mu,X^\nu \right]_- = - i \delta_\mu^\nu\,, \,\,\,\,\,\,\,\,
P_\mu|p\rangle = p_\mu |p\rangle\,, \,\,\,\,\,\,\,\,
\langle p | p' \rangle = \delta^4(p-p')\,,& \nonumber \\
&\int|p\rangle\langle p|dp = I\,, \,\,\,\,\,\,\,\,
\langle x |P_\mu| y \rangle = -i\partial_\mu\delta^4(x-y)\,,\,\,\,\,\,\,\,\,
\langle x | p \rangle = \frac{1}{(2\pi)^2}e^{ipx}\,,& \nonumber \\
&\left[\Pi_\mu,\Pi_\nu \right]_- = - igF_{\mu\nu}(X)\,, \,\,\,\,\,\,\,\,
\Pi_\mu = -P_\mu - g A_\mu(X)\,.& \nonumber
\end{eqnarray}
The equation (\ref{grennmod}) implies the formal solution for the operator
$ \tilde{S}^{\rm c}$:
$$
\tilde{S}^{\rm c} = \left[
\Pi_\nu\gamma^\nu -
m \gamma^5 -\frac{\mu}{2}\gamma^5\sigma^{\alpha\beta}F_{\alpha\beta}
\right]^{-1} \, .
$$
The operator in square brackets is a pure Fermi one, if one reckons
$\gamma$-matrices as Fermi operators. In general case
the inverse operator to a Fermi operator
$A$ can be presented by means of an integral over the super-proper time
($\lambda,\chi$)  of an
exponential with an even exponent \cite{gitfrad},
$$
A^{-1} = \int_0^\infty \, d\lambda
\int  e^{i(\lambda(A^2+i\epsilon)+\chi A)} d\,\chi\,,
$$
 where $\lambda$ is
an even and $\chi$ is an odd (Grassmann) variable, the latter
anticommutes with $A$ by definition.
In our case
\begin{eqnarray}
A &=& \Pi_\nu\gamma^\nu -
m \gamma^5 -
i\frac{\mu}{2}\gamma^5F_{\alpha\beta}\gamma^\alpha\gamma^\beta , \nonumber \\
A^2 &=& \Pi^2 - m^2 -
i \left(m\mu +\frac{g}{2} \right)F_{\alpha\beta}\gamma^\alpha\gamma^\beta
+ \frac{\mu^2}{4}\left(F_{\alpha\beta}
\gamma^\alpha\gamma^\beta\right)^2\ \nonumber \\
&& -  i\mu\gamma^5\gamma^\alpha \left[F_{\alpha\beta},\Pi^\beta\right]_+\,,
\nonumber
\end{eqnarray}
where was used
$\left[\Pi_\nu\gamma^\nu,i\frac{\mu}{2}
\gamma^5 F_{\alpha\beta}\gamma^\alpha\gamma^\beta\right]_+
=i\mu\gamma^5\gamma^\alpha \left[F_{\alpha\beta},\Pi^\beta\right]_+$.
So we get:
$$
\tilde{S}^{\rm c} =\int_0^\infty \, d\lambda
\int e^{-i\hat{\cal H}(\lambda,\chi)} d\chi\,\,,
$$
where:
\begin{eqnarray}
&&\hat{{\cal H}}(\lambda,\chi) = \lambda \left\{
m^2 - \Pi^2 + i\left(m\mu
+\frac{g}{2} \right)F_{\alpha\beta}  \gamma^\alpha\gamma^\beta
- \frac{\mu^2}{4}\left(
F_{\alpha\beta}\gamma^\alpha\gamma^\beta\right)^2
\right.  \nonumber \\
&&\left.
+ i\mu\gamma^5\gamma^\alpha \left[F_{\alpha\beta},\Pi^\beta\right]_+
\right\} + \left(
\Pi_\nu\gamma^\nu - m \gamma^5 -
i\frac{\mu}{2}\gamma^5
F_{\alpha\beta}\gamma^\alpha\gamma^\beta
\right)\chi\,.\nonumber
\end{eqnarray}
Thus, the Green's function (\ref{prop}) takes the form:
\begin{equation}
\tilde{S}^{\rm c}(x_{\rm out},x_{\rm in}) =\int_0^\infty \, d\lambda
\int \langle x_{\rm out} | e^{-i\hat{\cal H}(\lambda,\chi)}|x_{\rm in}
\rangle d\chi\,\,.
\label{prop1}
\end{equation}

Now we are going to present the matrix element entering in the expression
(\ref{prop1}) by means of a path integral. In spite of the operator
$\hat{{\cal H}}(\lambda,\chi)$ has the $\gamma$-matrix structure, one can
do this similar to the usual way, namely, first write
$\exp -i\hat{{\cal H}} = \left(\exp -i\hat{{\cal H}}/N \right)^N $
and then insert $(N-1)$
resolutions of identity $\int|x\rangle\langle x|dx = I$ between all the
operators $\exp -i\hat{{\cal H}}/N$. Besides, we introduce
$N$ additional integrations over $\lambda$ and $\chi$ to transform then the
ordinary integrals over these variables into the corresponding path-integrals:
\begin{eqnarray}
\label{paint}
&& \tilde{S}^{\rm c}(x_{\rm out},x_{\rm in}) =  \lim_{N\rightarrow \infty}
\int_0^\infty d\, \lambda_0
\int \prod_{k=1}^{N}\langle x_k |
e^{-i\hat{{\cal H}}(\lambda_k,\chi_k)\Delta \tau} | x_{k-1} \rangle \\
&&
\delta(\lambda_k-\lambda_{k-1})\delta(\chi_k-\chi_{k-1})
d\, \chi_0 d\, x_1 ... d\, x_{N-1}
d\, \lambda_1 ... d\, \lambda_N d\, \chi_1 ... d\, \chi_N\,, \nonumber
\end{eqnarray}
where $\Delta \tau = 1/N$, $x_0=x_{\rm in}$, $x_N=x_{\rm out}$.

\noindent
Bearing in mind the limiting process, one can, as usual, restrict himself by
calculation of the matrix elements from (\ref{paint}) approximately,
\begin{equation}
\langle x_k |
e^{-i\hat{{\cal H}}(\lambda_k,\chi_k)\Delta \tau} | x_{k-1} \rangle \approx
\langle x_k |
1{-i\hat{{\cal H}}(\lambda_k,\chi_k)\Delta \tau} | x_{k-1} \rangle,
\label{aprox}
\end{equation}
using the resolutions of identity $\int|p\rangle\langle p| d\,p$. In this
connection it is  important to notice that the operator
$\hat{{\cal H}}(\lambda_k,\chi_k)$ has the symmetric form in the operators
$\hat{x}$ and $\hat{p}$ by origin. Indeed, the only terms in
$\hat{{\cal H}}(\lambda_k,\chi_k)$, which contain products of these operators
are $[\hat{p}_\alpha,A^\alpha(\hat{x})]_+$ and
$[\hat{p}_\alpha,F^{\alpha\beta}(\hat{x})]_+$. One can verify that these are
maximal symmetrized expressions, which can be combined from entering operators
(see also remark in ref. \cite{witt}, p. 388). Thus, one can write
$$
\hat{{\cal H}}(\lambda,\chi) = {\rm Sym}_{(\hat{x},\hat{p})}\,\,
{\cal H}(\lambda,\chi,\hat{x},\hat{p}),
$$
where ${\cal H}(\lambda,\chi,{x},{p})$ is the Weyl symbol of the
operator $\hat{{\cal H}}(\lambda,\chi)$ in the sector of ${x}$, ${p}$,
\begin{eqnarray}
&&{\cal H}(\lambda,\chi,x,p) =  \lambda \left(m^2-{\cal P}^2
+i\left(m\mu+\frac{g}{2}\right)F_{\alpha\beta}  \gamma^\alpha\gamma^\beta
 + 2i\mu\gamma^5 F_{\alpha\beta}\gamma^\alpha
{\cal P}^\beta  \right. \nonumber \\
&&-\left. \frac{\mu^2}{4}(F_{\alpha\beta}\gamma^\alpha\gamma^\beta)^2 \right)
+\left(  {\cal P}_\nu\gamma^\nu - m\gamma^5 -
i\frac{\mu}{2}\gamma^5 F_{\alpha\beta}\gamma^\alpha\gamma^\beta
\right)\chi\,,\nonumber
\end{eqnarray}
where ${\cal P}_\nu= -p_\nu-gA_\nu(x)\,$.

\noindent
That is a general statement \cite{ber1}, which can be easily checked in that
concrete case by direct calculations, that the matrix elements (\ref{aprox})
are expressed in terms of
 the Weyl symbols in the middle point $\overline{x}_k = (x_k+x_{k-1})/2$.
Taking all that into account, one can see that in the limiting process the
matrix elements (\ref{aprox}) can be replaced by the expressions
\begin{equation}
\int \frac{d\,p_k}{(2\pi)^4}\exp i \left[
p_k\frac{x_k-x_{k-1}}{\Delta\tau} - {\cal H}(\lambda_k,\chi_k,
\overline{x}_k,p_k) \right]\Delta\tau\,,
\label{dez}
\end{equation}
which all are noncommuting  due to the $\gamma$-matrix structure and are
situated in (\ref{paint}) so that the numbers $k$ increase
from the right to the left. For the
two $\delta$-functions, accompanying each matrix element (\ref{aprox}) in the
expression (\ref{paint}), we use the integral representations
$$
\delta(\lambda_k-\lambda_{k-1})\delta(\chi_k-\chi_{k-1}) =
\frac{i}{2\pi} \int e^{i\left[
\pi_k\left(\lambda_k-\lambda_{k-1}\right)+
\nu_k\left(\chi_k   -\chi_{k-1}   \right)
\right]}d\, \pi_k d\, \nu_k,
$$
where $\nu_k$ are odd variables. Then we attribute formally to the
$\gamma$-matrices, entering into  (\ref{dez}), index $k$, and
then we attribute to all quantities the ``time'' $\tau_k$, according the
index $k$ they have, $\tau_k=\Delta\tau k$, so that $\tau \in [0,1]$.
Introducing the T-product, which acts on $\gamma$-matrices, it is possible
to gather all the
expressions, entering in  (\ref{paint}),
in one exponent and deal then with the $\gamma$-matrices
like with odd (Grassmann) variables.
Taking into account all was said, we get for the right side of (\ref{paint}):
\begin{eqnarray}
\label{pint}
&& \tilde{S}^{\rm c}(x_{\rm out},x_{\rm in}) =  \\
&& {\rm T}\int_0^\infty \, d\lambda_0
\int  \exp \left\{i\int_0^1 \left[ \lambda \left({\cal P}^2 - m^2-
i\left(m\mu +\frac{g}{2}\right)F_{\alpha\beta}  \gamma^\alpha\gamma^\beta
\right.\right.\right.\nonumber \\
&&  \left.
 - 2i\mu\gamma^5 F_{\alpha\beta}\gamma^\alpha {\cal P}^\beta  +
\frac{\mu^2}{4}(F_{\alpha\beta}\gamma^\alpha\gamma^\beta)^2
\right) +\left( m\gamma^5 - {\cal P}_\nu\gamma^\nu
+i\frac{\mu}{2}\gamma^5 F_{\alpha\beta}\gamma^\alpha\gamma^\beta
\right)\chi \nonumber \\
&&\left.\left.\rule{0ex}{3ex} + p\dot{x} + \pi\dot{\lambda} + \nu\dot{\chi}
\right]d\tau\right\} d \chi_0  {\cal D}x {\cal D}p {\cal D}\lambda {\cal D}\pi
 {\cal D}\chi {\cal D}\nu\,, \nonumber
\end{eqnarray}
where   $x$,  $p$,  $\lambda$,  $\pi$ ,
are even and $\chi$,
$\nu$, odd trajectories, obeying the boundary conditions
 $\,\,\,x(0)=x_{\rm in}$, $\,\,\,x(1)=x_{\rm out}$,
$\,\,\,\lambda (0) = \lambda_0$,
$\,\,\,\chi(0) = \chi_0$. The operation of T-ordering
acts on the $\gamma$-matrices, which suppose formally to
depend on time $\tau$.

The expression (\ref{pint}) can be reduced to:
\begin{eqnarray}
&& \tilde{S}^{\rm c}(x_{\rm out},x_{\rm in}) = \nonumber \\
&&\int_0^\infty \, d\lambda_0
\int  \exp \left\{
i\int_0^1 \left[
\lambda\left( {\cal P}^2 -m^2-i\left(m\mu
+\frac{g}{2} \right)F_{\alpha\beta}
\frac{\delta_l}{\delta \rho_\alpha}
\frac{\delta_l}{\delta \rho_\beta}
 \right.\right.\right.
\nonumber \\
&&\left.
- 2i\mu F_{\alpha\beta}
{\cal P}^\beta
\frac{\delta_l}{\delta \rho_5}
\frac{\delta_l}{\delta \rho_\alpha}
+ \frac{\mu^2}{4} \left( F_{\alpha\beta}
\frac{\delta_l}{\delta \rho_\alpha}
\frac{\delta_l}{\delta \rho_\beta}\right)^2
 \right)
+\left( m\frac{\delta_l}{\delta \rho_5}
   - {\cal P}_\nu\frac{\delta_l}{\delta \rho_\nu} \right. \nonumber \\
&&\left.\left.\left. + i\frac{\mu}{2} F_{\alpha\beta}
\frac{\delta_l}{\delta \rho_5}\frac{\delta_l}{\delta \rho_\alpha}
\frac{\delta_l}{\delta \rho_\beta}
\right)\chi  + p\dot{x} + \pi\dot{\lambda} + \nu\dot{\chi}
\right]d\tau\right\}
d \chi_0  {\cal D}x {\cal D}p {\cal D}\lambda {\cal D}\pi
 {\cal D}\chi {\cal D}\nu \nonumber \\
&&\left. \times
 {\rm T}\exp \int_0^1\rho_n(\tau)\gamma^n d\tau \right|_{\rho=0},\nonumber
\end{eqnarray}
where
 five Grassmannian sources $\rho_n(\tau)$ are
introduced, which  anticommute with the
$\gamma$-matrices by definition.

\noindent
Using the representation (\ref{a5}) for
$ {\rm T}\exp \int_0^1\rho_n(\tau)\gamma^n d\tau $ in the form of a
Grassmannian path-integral (see Appendix),
 we get the Hamiltonian path-integral representation
for the Green's function in question:
\newpage
\begin{eqnarray}
\label{hpint}
&& \tilde{S}^{\rm c}(x_{\rm out},x_{\rm in}) =  \\
&& \exp\left(i\gamma^n\frac{\partial_l}{\partial\theta^n} \right)
\int_0^\infty \, d\lambda_0 \int
\exp \left\{
i\int_0^1 \left[ \lambda\left(
{\cal P}^2 - 8i\mu\psi^5 F_{\alpha\beta}{\cal P}^\alpha\psi^\beta
\right.\right.\right. \nonumber \\
&&\left. + 2igF_{\alpha\beta}\psi^\alpha\psi^\beta - M^2 \right)
+2i\left(
{\cal P}_\alpha\psi^\alpha - M\psi^5\right)\chi
 -i\psi_n\dot{\psi}^n
+ p\dot{x} \nonumber \\
&&\left.\left.\left.\rule{0ex}{3ex} + \pi \dot{\lambda} +\nu \dot{\chi}
\right] d\tau
+ \psi_n(1)\psi^n(0) \right\}  d \chi_0 {\cal D}x
{\cal D}p {\cal D}\lambda {\cal D}\pi
 {\cal D}\chi {\cal D}\nu {\cal D}\psi \right|_{\theta=0}, \nonumber
\end{eqnarray}
where: ${\cal P}_\nu= -p_\nu-gA_\nu(x)\,$,
$M = m - 2i\mu F_{\alpha\beta}\psi^\alpha\psi^\beta$;
$x$,  $p$,  $\lambda$,  $\pi$, are even and $\chi$,
$\nu$, $\psi^n$,  odd trajectories, obeying the boundary conditions
 $x(0)=x_{\rm in}$, $x(1)=x_{\rm out}$, $\lambda (0) = \lambda_0$,
$\chi(0) = \chi_0\,$, $\psi^n(1)+\psi^n(0) = \theta^n$, and $\theta^n$ are
some Grassmannian variables.

Integrating over momenta $p$,
we come to the corresponding path-integral in the Lagrangian form:
\begin{eqnarray}
\label{pathin}
&& \tilde{S}^{\rm c}(x_{\rm out},x_{\rm in}) = \\
&&\exp\left(i\gamma^n\frac{\partial_l}{\partial\theta^n} \right)
\int_0^\infty \, de_0
\int    \exp \left\{
i\int_0^1 \left[  -\frac{\dot{x}^2}{2e} - e\frac{M^2}{2}
- \dot{x}^\alpha\left(gA_\alpha \right. \right. \right. \nonumber \\
&&\left.+4i\mu \psi^5F_{\alpha\beta}\psi^\beta
\right) +igeF_{\alpha\beta}\psi^\alpha\psi^\beta
+ i\left(\frac{\dot{x}_\alpha\psi^\alpha}{e}-M^*\psi^5
\right)\chi -i\psi_n\dot{\psi}^n \nonumber \\
&&\left.\left.\left.  + \pi \dot{e} +\nu \dot{\chi}
\right] d\tau + \psi_n(1)\psi^n(0) \right\}{\cal M}(e)d \chi_0
 {\cal D}x {\cal D}e {\cal D}\pi
 {\cal D}\chi {\cal D}\nu {\cal D}\psi \right|_{\theta=0},\nonumber
\end{eqnarray}
where $M^*=m+ 2i\mu F_{\alpha\beta}\psi^\alpha\psi^\beta$, and
 the measure ${\cal M}(e)$ has the form:
\begin{equation}
{\cal M}(e)=\int {\cal D}p \exp \left[
\frac{i}{2}\int_0^1 ep^2 d\,\tau
\right].
\label{meas}
\end{equation}
The discussion of the role of the
measure (\ref{meas}) can be found in \cite{gitfrad}.

\section{Action of Spinning Particle with
Anomalous Magnetic Momentum }

The exponent in the integrand (\ref{pathin}) can be considered as an
effective and nondegenerate Lagrangian action of a spinning particle with an
anomalous magnetic momentum.
It consists of two principal parts. The first one,
which unifies two summands with the
derivatives of $e$ and $\chi$, can be treated as a
gauge fixing term $S_{\rm GF}$,
$$
S_{\rm GF} = \int_0^1\left(
\pi \dot{e} + \nu \dot{\chi}
\right)d\,\tau,
$$
and corresponds, in fact, to the gauge conditions
\begin{equation}
\dot{e} = \dot{\chi} = 0.
\label{gc}
\end{equation}
The rest part of the effective action can be treated as a gauge invariant
action of a spinning particle with an anomalous magnetic momentum.
It has the form
\begin{eqnarray}
 S &=& \int_0^1 \left[
 -\frac{\dot{x}^2}{2e}  - e\frac{M^2}{2} - \dot{x}^\alpha\left(gA_\alpha
+4i\mu\psi^5 F_{\alpha\beta}\psi^\beta \right)
+igeF_{\alpha\beta}\psi^\alpha\psi^\beta
  \right.\nonumber \\
&&\left. +i\left(\frac{\dot{x}_\alpha\psi^\alpha}{e} -
M^*\psi^5
\right)\chi -i\psi_n\dot{\psi}^n
\right]d\,\tau\,\,.
\label{lagr}
\end{eqnarray}
Indeed, one can verify that there are two types of gauge transformations, under
which the actions is invariant, in accordance with two gauge
conditions (\ref{gc}).
The first one coincide with reparametrizations (\ref{re}) and the
second one is a generalization of supertransformations (\ref{su}),
\begin{eqnarray}
&&\delta x = i\psi\epsilon \,\,,\,\,\,\,\,\,
\delta e = i\chi \epsilon \,\,,\,\,\,\,\,\,
\delta \chi = \dot{\epsilon} \,\,,\,\,\,\,\,\,
\delta\psi^\alpha = \frac{1}{2e}(\dot{x}^\alpha
+ i\chi\psi^\alpha)\epsilon \,\, ,  \nonumber \\
&&\delta\psi^5 = \left[\frac{M}{2}
-\frac{i}{me}\psi^5\left(
\dot{\psi}^5 - 2\mu F_{\alpha\beta}\dot{x}^\alpha\psi^\beta
-\frac{M^*}{2}\chi
\right)
\right]\epsilon\,\,.
\label{sup}
\end{eqnarray}

Let us analyze the theory with the action (\ref{lagr}).
The Lagrangian equations of motion have the form:
\begin{eqnarray}
\label{eq1}
\frac{\delta S}{\delta e} &=& \frac{1}{e^2}\left(\frac{\dot{x}^2}{2}
-i\dot{x}_\alpha\psi^\alpha\chi\right) - \frac{M^2}{2}
+ igF_{\alpha\beta}\psi^\alpha\psi^\beta = 0\,,  \\
\label{eq2}
\frac{\delta_r S}{\delta\chi} &=& i\left(
\frac{\dot{x}_\alpha}{e}\psi^\alpha - M^*\psi^5\right)=0\,,\\
\label{eq3}
\frac{\delta_r S}{\delta\psi^\alpha} &=& 2i\dot{\psi}_\alpha + 2ie\left(
g+2M\mu\right)F_{\beta\alpha}\psi^\beta-4\mu\psi^5\left(
i\dot{x}^\beta - \chi\psi^\beta\right)F_{\beta\alpha}  \nonumber \\
&&-i\frac{\dot{x}_\alpha}{e}\chi = 0\,,  \\
\label{eq4}
\frac{\delta_r S}{\delta\psi^5} &=& -2i\dot{\psi}^5 + 4i\mu \dot{x}^\alpha
F_{\alpha\beta}\psi^\beta + iM^*\chi = 0\,, \\
\label{eq5}
\frac{\delta S}{\delta x^\alpha} &=&
\frac{d}{d\tau}\left(\frac{\dot{x}_\alpha}{e}\right)
+ g\dot{x}^\beta F_{\beta\alpha}
- 4i\mu\psi^5\dot{x}^\beta\psi^\gamma F_{\beta\alpha,\gamma}
+ ie \left(g+2M\mu \right)
F_{\beta\gamma,\alpha}\psi^\beta\psi^\gamma\nonumber \\
&&+ 2\mu\psi^5\chi F_{\beta\gamma,\alpha}\psi^\beta\psi^\gamma = 0 \,.
\end{eqnarray}
Because of the existence of the
gauge transformations (\ref{re}) and (\ref{sup})
there are two  identities between all these equations. That means that the
number of independent equations is less
than the number of the variables, and we can fix
two of the latter
 by imposing of some gauge conditions. One can choose (it is also
seen from the Hamiltonian analysis which follows) the gauge conditions
 $\chi=0$ and $e=1/m$ to simplify the analysis of the equations
(\ref{eq1}-\ref{eq5}).
Our aim is to show that equations (\ref{eq1}-\ref{eq5}) describe a
spinning particle with anomalous momentum $\mu$. To this end we need to use
the nonrelativistic approximation and consider the case of a weak magnetic
field. In such a case $F_{0i} = 0$, and
$F_{ij} = -\epsilon_{ijk}B_k$,
where $B_k$ are components of a
magnetic field $\vec{B}$ and $\epsilon_{ijk}$ is
the tree dimensional Levi-Civita symbol. It is easy to see from the equations
(\ref{eq1},\ref{eq2},\ref{eq4}) that in the  gauge and in the
above mentioned approximation we have:
\begin{equation}
\dot{x}^0 \approx 1 \,,\,\,\,\,\,\,
\dot{x}^i \approx v^i = \frac{dx^i}{dx^0}\,,
\label{aaaa}
\end{equation}
 Introducing the three
dimensional spin vector \cite{ber}
$$
s_k= i\epsilon_{kjl}\psi_l\psi_j \,,
$$
and using (\ref{aaaa}), we get from the equations (\ref{eq3}) and (\ref{eq5}):
\begin{eqnarray}
\dot{\vec{s}} &=&
2\left(\frac{g}{2m}+\mu\right)\vec{s}\times\vec{B}\,,\nonumber \\
m\dot{\vec{v}} &=& g\vec{v}\times\vec{B} +
2\left(\frac{g}{2m}+\mu\right)\nabla\left(\vec{s}\cdot\vec{B}\right)\,,
\label{eqmot2}
\end{eqnarray}
The equations (\ref{eqmot2}) describe \cite{pan}
a nonrelativistic motion of a particle
with total spin momentum $\vec{s}$, and total magnetic momentum
 $2\left(\frac{g}{2m}+\mu\right)\vec{s}$. Taking into account that the
giromagnetic ratio for electron equals 2, one can conclude that $\mu$
corresponds to the anomalous magnetic momentum of electron, what
confirms the interpretation of the action (\ref{lagr}).

 Going over to the
Hamiltonian formalism, we introduce the canonical momenta:
\begin{eqnarray}
&&p_\alpha = \frac{\partial L}{\partial\dot{x}^\alpha} = -\frac{1}{e}\left(
\dot{x}_\alpha - i\psi_\alpha\chi\right) - gA_\alpha
- 4i\mu\psi^5 F_{\alpha\beta}\psi^\beta, \nonumber \\
&&P_e = \frac{\partial L}{\partial\dot{e}} = 0, \,\,\,\,\,\,\,\,
P_\chi = \frac{\partial_r L}{\partial\dot{\chi}} = 0, \,\,\,\,\,\,\,\,
P_n =  \frac{\partial_r L}{\partial\dot{\psi}^n} = -i\psi_n\,\,\,.
\label{momenta}
\end{eqnarray}
It follows from the
equation  (\ref{momenta}) that there exist primary constraints \break
 $\Phi_A^{(1)}=0$,
\begin{equation}
\Phi_A^{(1)}=\left\{
\begin{array}{l}
\Phi_1^{(1)}= P_\chi\,\,\,, \\
\Phi_2^{(1)}= P_e\,\,\,, \\
\Phi_{3n}^{(1)}= P_n +i\psi_n\,\,\,.
\end{array}
\right.
\label{primary}
\end{equation}
We construct the  Hamiltonian $H^{(1)}$, according to the standard procedure
(we use the notations of the book \cite{quan}),
$$
H^{(1)}= H + \lambda_A \Phi_A^{(1)},\,\,\, {\rm where}\,\,\,\,
H = \left.\left(
\frac{\partial_rL}{\partial\dot{q}}\dot{q}-L\right)
\right|_{\frac{\partial_rL}{\partial\dot{q}}=P},
\,\,\,\,\,\, q = (x,e,\chi,\psi^n)\,\,\,,
$$
and get for $H$:
\begin{eqnarray}
H = &-&\frac{e}{2} \left( {\cal P}^2 - 8i\mu
\psi^5F_{\alpha\beta}{\cal P}^\alpha\psi^\beta
+2igF_{\alpha\beta}\psi^\alpha\psi^\beta -  M^2
\right) \nonumber \\
&-&i\left(  {\cal P}_\alpha\psi^\alpha -M\psi^5 \right)\chi\, .\nonumber
\end{eqnarray}
{}From the
 conditions of the conservation of the primary constraints $\Phi_{1,2}^{(1)}$
in time $\tau$ ,
$\dot{\Phi}^{(1)}_{1,2} = \left\{{\Phi}^{(1)}_{1,2},H^{(1)} \right\} = 0$,
we find the secondary constraints $\Phi_{1,2}^{(2)} = 0$,
\begin{eqnarray}
\label{first}
\Phi_1^{(2)}&=& {\cal P}_\alpha\psi^\alpha -M\psi^5 =0,  \\
\Phi_2^{(2)}&=&
{\cal P}^2 - 8i\mu \psi^5F_{\alpha\beta}{\cal P}^\alpha\psi^\beta
+2igF_{\alpha\beta}\psi^\alpha\psi^\beta -  M^2=0,
\label{second}
\end{eqnarray}
and  the same conditions for the constraints $\Phi^{(1)}_{3n}$ give
equations for the determination of $\lambda_{3n}$.
Thus, the Hamiltonian $H$ appears to be
 proportional to constraints, as one can
expect in the case of a re\-pa\-ra\-me\-tri\-za\-tion invariant theory,
$$
H =  i\chi\Phi^{(2)}_1 -\frac{e}{2} \Phi^{(2)}_2 .
$$
No more secondary constraints  arise from the Dirac procedure, and the
Lagrange's multipliers $\lambda_{1}$ and $\lambda_{2}$ remain undetermined,
in perfect correspondence with the fact that the number of gauge
transformations parameters equals two for the theory in question \cite{quan}.
One can go over from the initial set of constraints
$\left(\Phi^{(1)},\Phi^{(2)}\right)$ to  the equivalent one
$ \left(\Phi^{(1)},T\right),$ where:
\begin{equation}
 T = \Phi^{(2)} +
\frac{i}{2} \frac{\partial_r\Phi}{\partial\psi^n}^{(2)}\Phi^{(1)}_{3n}\,\,.
\label{dt}
\end{equation}
The new set of constraints can be explicitly divided in
a set of the first-class constraints, which is
$\left(\Phi^{(1)}_{1,2},T\right)$ and in a set of the
second-class constraints,
 which is $\Phi^{(1)}_{3n}$.

\section{Quantization}

First we consider an operator quantization, expecting to get in this procedure
the Dirac-Pauli equation (\ref{gendirac}).

In our case  we perform only a partial gauge fixing, by imposing
 the supplementary gauge conditions
$\Phi^{\rm G}_{1,2}=0$ to the primary first-class constraints
$\Phi^{(1)}_{1,2}\,\,$,
\begin{equation}
\Phi^{\rm G}_1 = \chi=0, \,\,\,\,\,\,
\Phi^{\rm G}_2 = e = 1/m\,,
\label{gauge}
\end{equation}
which coincide with those we used in the
Lagrangian analysis. One can check that
the conditions of the conservation in time of the supplementary constraints
(\ref{gauge}) give equations for determination of the
multipliers $\lambda_1$ and $\lambda_2$.
Thus, on this stage we reduced our Hamiltonian theory to one
with the first-class constraints $T$ and second-class ones
$\varphi = \left(\Phi^{(1)},\Phi^{\rm G} \right)$.
After that
 we will use the so called Dirac method for
 systems with first-class constraints \cite{yesh}, which, being
generalized to the presence of second-class constraints, can be
formulated as follow: the commutation relations between operators are
calculated according to the Dirac brackets with respect to the
second-class  constraints only; second-class constraints operators
equal zero; first-class constraints as operators are not zero, but,
are considered in sense of restrictions on state vectors.
All the operator equations have to be realized in some Hilbert space.

The sub-set of the second-class constraints
$\left(\Phi^{(1)}_{1,2},\Phi^{\rm G}\right)$ has a special form \cite{quan},
so that one can use it for eliminating of the variables $e,P_e,\chi,P_\chi$,
 from the consideration, then,
for the rest of the variables $x,p,\psi^n$,
the Dirac brackets with respect to the constraints $\varphi$
reduce to ones with respect to the constraints $\Phi^{(1)}_{3n}$ only and
can be easy calculated,
$$
 \left\{x^\alpha,p_\beta \right\}_{D(\Phi^{(1)}_{3n})} =
 \delta^\alpha_\beta \,,\,\,\,\,\,\,\,
 \left\{\psi^n,\psi^m \right\}_{D(\Phi^{(1)}_{3n})} =
\frac{i}{2}\eta^{nm}\,,
$$
while  others Dirac brackets vanish.
Thus, the commutation relations for the operators
$\hat{x},\hat{p},\hat{\psi}^n$, which correspond to the variables
$x,p,\psi^n$ respectively, are
\begin{eqnarray}
 \left[\hat{x}^\alpha,\hat{p}_\beta \right]_-
&=& i\left\{x^\alpha,p_\beta \right\}_{D(\Phi^{(1)}_{3n})} =
\delta^\alpha_\beta\,, \nonumber \\
 \left[\hat{\psi}^m,\hat{\psi}^n \right]_+
&=& i\left\{\psi^m,\psi^n \right\}_{D(\Phi^{(1)}_{3n})}=
-\frac{1}{2}\eta^{mn}.
\label{comm}
\end{eqnarray}
Besides, the operator equations hold:
\begin{equation}
\hat{\Phi}^{(1)}_{3n}=  \hat{P}_n + i \hat{\psi}_n =0 .
\label{oe}
\end{equation}
The commutation relations (\ref{comm}) and the equations
 (\ref{oe}) can be realized
in a space of  four columns $\Psi(x)$ dependent on $x^\alpha$
as: $\hat{x}^\alpha$ are
operators of multiplication, $\hat{p}_\alpha = -i\partial_\alpha$,
$\hat{\psi}^\alpha = \frac{i}{2}\gamma^5\gamma^\alpha$, and
$\hat{\psi}^5 = \frac{i}{2}\gamma^5$,
where $\gamma^n$ are the $\gamma$-matrices
$(\gamma^\alpha,\gamma^5)$.
The first-class constraints $\hat{T}$ as operators have to annihilate
physical vectors; in virtue of (\ref{oe}), (\ref{dt}) these conditions
 reduce to the equations:
\begin{equation}
\hat{\Phi}^{(2)}_{1,2}\Psi(x)=0,
\label{t}
\end{equation}
where $\hat{\Phi}^{(2)}_{1,2}$ are operators, which correspond to the
constraints (\ref{first}), (\ref{second}). There is no ambiguity in the
construction of the operator $\hat{\Phi}^{(2)}_1$, according to the classical
function $\Phi^{(2)}_1$ from (\ref{first}). Thus, taking into account
the realizations of the commutation relations (\ref{comm}), one easily can see
that the first equation  (\ref{t})
reproduces the Dirac-Pauli equation  (\ref{gendirac}).
As to the construction of the operator $\hat{\Phi}^{(2)}_2$, according to the
classical function ${\Phi}^{(2)}_2$ from (\ref{second}), we meet here an
ordering problem since the constraint ${\Phi}^{(2)}_2$ contains terms
with products of the momenta and functions of the coordinates, namely terms
of the form $p_\alpha A^\alpha$,
$p_\alpha F^{\alpha\beta}$.
For such  terms we choose the symmetrized form of the
corresponding operators,
\begin{equation}
p_\alpha A^\alpha \rightarrow \frac{1}{2}
\left[\hat{p}_\alpha,A^\alpha(\hat{x}) \right]_{+} \, , \,\,\,\,\,\,\,\,
p_\alpha F^{\alpha\beta} \rightarrow \frac{1}{2}
\left[\hat{p}_\alpha,F^{\alpha\beta}(\hat{x}) \right]_{+} \, ,
\label{real}
\end{equation}
which, in particular, provides the hermiticity of the operator
$\hat{\Phi}^{(2)}_2$. But the main reason is, the correspondence rule
(\ref{real}) provides the consistency of the two equations (\ref{t}). Indeed,
in this case we have
\begin{equation}
\hat{\Phi}^{(2)}_2 = \left(\hat{\Phi}^{(2)}_1 \right)^2,
\label{sq}
\end{equation}
and the second equation  (\ref{t}) appears to be merely the consequence of the
first equation  (\ref{t}), i.e. of the Dirac-Pauli equation  (\ref{gendirac}).
To verify the validity of (\ref{sq}), one needs only to take into
 account that the operator, which corresponds to the term
$8 i\mu \psi^5F_{\alpha\beta}{\cal P}^\alpha\psi^\beta$
in the constraint
${\Phi}^{(2)}_2$ (\ref{second}), in virtue of the structure of the
$\gamma$-matrices, can be written in the form:
\begin{eqnarray}
&& 8 i\mu \psi^5F_{\alpha\beta}{\cal P}^\alpha\psi^\beta \rightarrow
i\mu\left[{F}_{\alpha\beta}(\hat{x}),\hat{{\cal P}}^\alpha
\right]_{+}\gamma^\beta =  \nonumber \\
&& 2i\mu F_{\alpha\beta}(\hat{x})\hat{\cal P}^\alpha\gamma^\beta +
\mu\partial^\alpha F_{\alpha\beta}(\hat{x})\gamma^\beta =
\left[\hat{\cal P}_\alpha\gamma^\alpha, \frac{\mu}{2}
\sigma^{\alpha\beta}F_{\alpha\beta}(\hat{x}) \right]_-\nonumber
\end{eqnarray}
To complete the operator
quantization, one has to present an inner product in the
space of realization of commutation relations. The general method of its
construction, in the frame of the Dirac method  we used,
is unfortunately still unknown. Nevertheless, in this concrete case,
the space of physical vectors, obeying the condition
(\ref{t}), can be transformed into a Hilbert space, if one
takes for the inner product ordinary  scalar product of solutions of
the Dirac equation, which does not depend on $x^0$, in spite of the
integration is fulfilled in it over $x^i$ only.
It is not
difficult to verify that the introduced operators, obey of natural properties
of hermiticity, which are known from the Dirac relativistic mechanics. In
particular, the operator $\hat{p}_0$, which has to be considered on the same
 foot with $\hat{p}_i$, is also hermitian on the solutions of the
Dirac-Pauli equation (\ref{gendirac}), in virtue of the above mentioned
independence of the scalar product on $x^0$.

Thus, we see that the operator quantization of the action (\ref{lagr})
reproduces the Dirac-Pauli quantum theory. To make the picture complete, one
oughts to discuss the path-integral quantization of the theory with the action
(\ref{lagr}). In fact, the problem reduces to a demonstration that the
path-integral  (\ref{hpint}) for the causal Green's function,
which is the propagator for the second quantized theory, can be interpreted
in the frame of the known formal methods of quantization of
gauge theories. However, a peculiarity of the theory in question
is that the Hamiltonian equals zero on
the constraints surface; we have avoided this problem in the operator
quantization, choosing the special Dirac method (direct ways of the
problem solving in the operator quantization were considered in ref.
[\citen{gittyu,gitgra}]).
Nevertheless, the path-integral (\ref{hpint})
can be interpreted in the frame of the
so called generalized Hamiltonian quantization (BFV method \cite{fra}).
 One can demonstrate that the exponent in path integral (\ref{hpint})
appears to be a generalized Hamiltonian action of the BFV method in a
special gauge.

We start the BFV consideration with the Hamiltonian formulation of our theory
after the partial gauge fixing (\ref{gauge}). In this case we have two
fist-class constraints (\ref{dt}) and the set of second class constraints
$\Phi^{(1)}_{3n}$ (\ref{primary}). We introduce new  canonical pairs:
odd $(\lambda_1,\pi_1)$, and even ghosts
$(\theta_1,\overline{\theta}_1)$,
$$
\theta_1 = \left(
\begin{array}{c}
c_1 \\ B_1
\end{array}
\right), \quad
\overline{\theta}_1 = \left(
\begin{array}{c}
 \overline{B}_1 \\ \overline{c}_1
\end{array}
\right),
$$
for the first-class constraint $T_1$;
even  $(\lambda_2,\pi_2)$, and odd ghosts $(\theta_2,\overline{\theta}_2)$,
$$
\theta_2 = \left(
\begin{array}{c}
c_2 \\ B_2
\end{array}
\right), \quad
\overline{\theta}_2 = \left(
\begin{array}{c}
 \overline{B}_2 \\ \overline{c}_2
\end{array}
\right),
$$
for the first-class constraint $T_2$
(${\rm gh}\,\theta = -{\rm gh}\,\overline{\theta} = 1$). After that we
construct
the fermionic generating charge $\Omega$ (also called BRST charge)
as a solution of the equation
$$
\left\{\Omega,\Omega \right\}_{D(\Phi^{(1)}_{3n})} = 0,
$$
obeying the conditions
\begin{eqnarray}
&&\Omega = \Omega_a\theta_a \,, \,\,\,\,\,\, {\rm gh}\,\Omega_a = 0 \,,
 \nonumber \\
&&\left.\Omega_a\right|_{\theta=\overline{\theta}=0} = \left(
\begin{array}{c}
T_a \\ \pi_a
\end{array}
\right)\,,
\,\,\,\,\, a=1,2. \nonumber
\end{eqnarray}
One can see that such a solution  has the form
\begin{equation}
\Omega=T_ac_a  + \pi_aB_a
+\frac{i}{4}\overline{B}_2(c_1)^2 .
\end{equation}
In the case of consideration,  the Hamiltonian on the constraints surface
equals zero, and one can write the BVF Hamiltonian in the form
$$
H_{\rm BFV} = \left\{
\Omega, \Psi
\right\}_{D(\Phi^{(1)}_{3n})}\,,
$$
where $\Psi$ is a gauge fermion,
\begin{eqnarray}
&&\Psi = \Psi_a\overline{\theta}_a \,, \,\,\,\,\,
{\rm gh}\, \Psi_a = 0 \,,  \nonumber \\
&& \left.\Psi_a\right|_{\theta=\overline{\theta}=0}=\chi_a \,, \,\,\,\,\,
a=1,2. \nonumber
\end{eqnarray}
Columns $\chi_a$ play
 the role of gauge conditions. To reproduce the path-integral (\ref{hpint}),
one needs to choose:
\begin{equation}
\chi_a = \left(\begin{array}{c}
\lambda_a \\ 0
\end{array} \right).
\end{equation}
So we get:
\begin{equation}
\Psi = \lambda_a\overline{B}_a ,
\end{equation}
which leads to:
\begin{equation}
H_{\rm BFV}= T_a\lambda_a
+\overline{B}_aB_a+\frac{i}{2}\overline{B}_2c_1\lambda_1\,,
\end{equation}
so that the correspondent BFV action has the form:
\begin{equation}
S_{\rm BFV}=\int d\tau \left( p\dot{x} + P_n\dot{\psi}^n
+ \pi_a\dot{\lambda}_a +
 \overline{\theta}_a \dot{\theta}_a
 -T_a\lambda_a -\overline{B}_aB_a-\frac{i}{2}\overline{B}_2c_1\lambda_1
\right)  \,.
\label{BFVaction}
\end{equation}
To construct the path-integral in BFV formulation one needs to use the well
known formula for a path-integral of theories with second-class constraints
\cite{fad} , which are in this case $\Phi^{(1)}_{3n}$.
Such an integral includes the factor
$$
\delta\left(\Phi^{(1)}_{3n} \right)
{\rm Sdet}^{1/2}\left\{\Phi^{(1)}_{3a},\Phi^{(1)}_{3b} \right\}\,.
$$
In the case of consideration the superdeterminant reduces to a
constant and the $\delta$-function lifts the integration over the
momenta $P_n$ and allows one to replace the constraints $T$ by
$\Phi^{(2)}$.
The trivial interaction of the
ghosts with $\lambda_1$ can be eliminated by means
of a shift of the variable $\overline{B}_1$, then the ghosts path-integral
separates.
Thus, after the renotations $\lambda_1\rightarrow -2i\chi$,
$\lambda_2\rightarrow -\lambda$,$\pi_1\rightarrow\frac{i}{2}\nu$,
$\pi_2\rightarrow -\pi$,
we can recognize in the path-integral (\ref{hpint}) the BFV
path-integral exponent. The operator factor in (\ref{hpint}) and the additional
integrations over $\lambda_0$, $\chi_0$ are connected with the special
choice of the Green's function of the Dirac-Pauli equation.

\section{Conclusion}

In the conclusion we would like to underline two points. First, that the
relationship between the model proposed and the one of spinning particle
reproduces on the classical level the relationship between the Dirac-Pauli
equation and Dirac equation. Second, the quantization of the model appears
to be rather instructive because of the ordering of operators plays here an
important role. In this sense the problem is analogous to the well known
problem of the quantization of a particle in curved space \cite{witt}.
It is remarkable that
 here the Weyl rule follows uniquely from the
condition of consistency of the Dirac quantization of theories with
first-class constraints. In this connection one can remind the result of the
work \cite{bat} where it was proved that only the Weyl ordering of operators
provides the correspondence principle in the Dirac quantization of theories
with second-class constraints.

\section*{Acknowledgments}

The authors would like to thank Professor Josif Frenkel for discussions and
one of them (AVS) thanks FAPESP(Brazil) for support.

\section*{Appendix}

Here we present the quantity
\begin{equation}
{\rm T}\exp \int_0^1 \rho_n(\tau)\gamma^n d\,\tau
\label{a1}
\end{equation}
by a Grassmannian path-integral. Remind that $\rho_n(\tau)$ are
odd variables; T-product acts on $\gamma$-matrices,
$[\gamma^m\gamma^n]_+ = 2\eta^{mn}$, $n=\overline{0,3},5$,
which suppose formally to
depend on time $\tau$, they anticommute by definition with  $\rho_n(\tau)$
and are considered in T-ordering procedure as Fermi operators

First we present (\ref{a1}) in the Sym-form. To this end we use a
formula, which is a version of the Weak theorem (
see, for example, reference \cite{vasi}). Let $\hat{\phi}(\tau)$
be some operators and $F(\hat{\phi})$ some functional on them. Then
$$
{\rm T}F(\hat{\phi}) = \left. {\rm Sym} \exp \left[
\frac{1}{2}\int\frac{\delta_r}{\delta\phi(\tau_1)}K(\tau_1,\tau_2)
\frac{\delta_r}{\delta\phi(\tau_2)}d\tau_1 d\tau_2
\right]F(\hat{\phi})\right|_{\phi=\hat{\phi}},
$$
where
$$
K(\tau_1,\tau_2) = {\rm T}\hat{\phi}(\tau_1)\hat{\phi}(\tau_2) -
{\rm Sym}\, \hat{\phi}(\tau_1)\hat{\phi}(\tau_2) =
\frac{1}{2} \epsilon(\tau_1-\tau_2)\left[
\hat{\phi}(\tau_1),\hat{\phi}(\tau_2)\right]_-,
$$
and
$\epsilon(\tau) = {\rm sgn} \,\tau\,$.
In the case of consideration,
$F(\hat{\phi})=\exp \int_0^1\hat{\phi}(\tau) d\tau  \,$,
$\hat{\phi}(\tau)=\rho_n(\tau)\gamma^n\,$, and
$K(\tau_1,\tau_2)=-\epsilon(\tau_1-\tau_2)\rho_n(\tau_1)\rho^n(\tau_2)$.
Consequently,
\begin{eqnarray}
&&{\rm T}\exp\int_0^1\rho_n(\tau)\gamma^n d\tau =  \nonumber\\
&&\exp\left[-\frac{1}{2}\int_0^1 d\tau_1\int_0^1 d\tau_2 \,
 \rho_n(\tau_1)\epsilon(\tau_1-\tau_2)\rho^n(\tau_2) \right]
\exp \int_0^1\rho_n(\tau)\gamma^n  \,.
\label{a2}
\end{eqnarray}
We have omitted the symbol Sym in the right side of (\ref{a2}) because of
the concrete structure of the functional.
The quadratic exponential from the right side of (\ref{a2}) can be presented
by means of a Gaussian path-integral over Grassmannian trajectories
$\xi^n(\tau)$,
\begin{eqnarray}
&&\exp\left[-\frac{1}{2}\int_0^1d\tau_1\int_0^1d\tau_2\,
\rho_n\left(\tau_1\right)\epsilon\left(\tau_1-\tau_2\right)
\rho^n\left(\tau_2\right) \right] = \nonumber \\
&&=\int\exp\left\{ \int_0^1\left[
 \frac{1}{4}\xi_n(\tau)\dot{\xi}^n(\tau) - i \rho_n(\tau)\xi^n(\tau)
\right]d\tau\right\}{\cal D}\xi \,,
\label{a3}
\end{eqnarray}
where a normalization factor is included in the measure of integration, so
that the integral in the right side of (\ref{a3}) equals unit at $\rho=0$.
The trajectories $\xi^n(\tau)$ anticommute with $\gamma$-matrices by definition
and obey the boundary conditions
\begin{equation}
\xi^n(0) + \xi^n(1) = 0,
\end{equation}
to make the path-integral in (\ref{a3}) invariant under the shifts
of integration variables.
One can also check \cite{buch} that the representation is available
\begin{equation}
\exp \left[
\int_0^1 \rho_n(\tau)\gamma^n d\tau\right] =
\left.\exp \left(i\gamma^n\frac{\partial_l}{\partial \theta^n} \right)
\exp \left[-i\int_0^1\rho_n(\tau)\theta^n d\tau \right|_{\theta=0}\right],
\label{a4}
\end{equation}
where $\theta^n$ are odd variables anticommuting with $\gamma$-matrices.
Gathering (\ref{a3}-\ref{a4}) and making then the change of variables
$$
\xi^n(\tau) + \theta^n = 2 \psi^n(\tau),
$$
we get the path integral representation for (\ref{a1})
\begin{eqnarray}
\label{a5}
&&{\rm T}\exp \int_0^1\rho_n(\tau)\gamma^n d\tau  =  \\
&&\left.\exp\left(i\gamma^n\frac{\partial_l}{\partial\theta^n}   \right)
\int\exp \left[
\int_0^1 \left(
\psi_n\dot{\psi}^n - 2i\rho_n\psi^n
\right) d\tau + \psi_n(1)\psi^n(0)
\right]{\cal D}\psi\right|_{\theta=0}. \nonumber
\end{eqnarray}

\end{document}